\documentclass[]{PoS}
\usepackage[authoryear,square]{natbib}
\bibpunct{(}{)}{;}{a}{}{,}

\title{Understanding the Neutron Star Population with the SKA}
\ShortTitle{Neutron Star Population with SKA}

\author{Thomas M. Tauris$^{1,2}$,
        \speaker{Victoria M. Kaspi}$\,^3$,
        Rene P. Breton$^4$, 
        Adam T. Deller$^5$,
        Evan~F.~Keane$^{6,7}$,
        Michael Kramer$^2$,
        Duncan R. Lorimer$^8$,
        Maura A. McLaughlin$^8$,
        Andrea Possenti$^9$,
        Paul S. Ray$^{10}$,
        Ben W. Stappers$^{11}$,
        Patrick Weltevrede$^{11}$
        \\ 
        $^1$Argelander-Institute for Astronomy, University of Bonn, Germany;\\
        $^2$Max-Planck Institute for Radio Astronomy, Bonn, Germany;\\
	$^3$Department of Physics, McGill University, Montreal, Canada;\\
        $^4$School of Physics and Astronomy, University of Southampton, UK;\\
        $^5$The Netherlands Institute for Radio Astronomy (ASTRON), Dwingeloo, The Netherlands;\\
        $^6$Centre for Astrophysics and Supercomputing, Swinburne University of Technology, Australia;\\
        $^7$ARC Centre of Excellence for All-Sky Astrophysics (CAASTRO);\\
        $^8$Department of Physics and Astronomy, West Virginia University, Morgantown, WV, USA;\\
	$^9$INAF-Osservatorio Astronomico di Cagliari, Selargius, Italy;\\
        $^{10}$Space Science Division, Naval Research Laboratory, Washington, DC, USA;\\
        $^{11}$School of Physics and Astronomy, The University of Manchester, UK 
        \\
        \\
        E-mail: \email{tauris@astro.uni-bonn.de}$\quad$ \email{vkaspi@physics.mcgill.ca}}

\abstract{
Since their discovery in the late 1960's the population of known neutron stars has grown to $\sim\!2500$.\\
The last five decades of observations have yielded many surprises and demonstrated that the 
observational properties of neutron stars are remarkably diverse. 
The surveys that will be performed with SKA (the Square Kilometre Array) will produce a further tenfold increase in the number of Galactic neutron stars known. 
Moreover, the SKA's broad spectral coverage, sub-arraying and multi-beaming capabilities will allow us to characterise these sources 
with unprecedented efficiency, in turn enabling a giant leap in the understanding of their properties. 
Here we review the neutron star population and outline our strategies for studying each of the growing number of diverse classes that are populating 
the ``neutron star zoo''. Some of the main scientific questions that will be addressed by the much larger 
statistical samples and vastly improved timing efficiency provided by SKA include: 
(i)    the spin period and spin-down rate distributions (and thus magnetic fields) at birth, 
       and the associated information about the supernovae wherein they are formed; 
(ii)   the radio pulsar--magnetar connection;
(iii)  the link between normal radio pulsars, intermittent pulsars and rotating radio transients;
(iv)    the slowest possible spin period for a radio pulsar (revealing the conditions at the pulsar death-line); 
(v)    proper motions of pulsars (revealing supernova kick physics); 
(vi)   the mass distribution of neutron stars; 
(vii)  the fastest possible spin period for a recycled pulsar 
       (constraining magnetosphere-accretion disc interactions, gravitational wave radiation and the equation-of-state);
(viii) the origin of high eccentricity millisecond pulsars; 
(ix)   the formation channels for recently identified triple systems; and finally
(x)    how isolated millisecond pulsars are formed.
As well as this lengthy (but not exhaustive) scientific shopping list, we can expect that 
the first phase of the SKA (SKA1), and in particular the full SKA (SKA2),  
will break new ground unveiling exotic and heretofore unknown systems that will challenge our current knowledge and theories, 
thus fostering the development of new research areas. 
Some possibilities for future landmark discoveries representing significant milestones in the astrophysics of 
compact objects include: 
(i)   sub-millisecond pulsars; 
(ii)  neutron stars born as millisecond pulsars; 
(iii) neutron stars with masses below 1.1 or above $2.5\;M_{\odot}$; 
(iv)  neutron star-black hole binaries; and 
(v)   a triple system containing a pair of neutron stars. 
}

\FullConference{
Advancing Astrophysics with the Square Kilometre Array\\
June 8-13, 2014\\
Giardini Naxos, Italy}

\newcommand{\skipthis}[1]{}
\begin{document}
\makeatletter
\setbox\@firstaubox\hbox{\small Thomas M. Tauris et al.}
\makeatother

\section{Introduction}\label{sec:intro}

The recent era of multi-wavelength observations has revealed a greater variety
of possibly distinct observational classes of neutron stars (NSs) than ever before.
In addition to isolated NSs, these compact objects
are found in binaries and even triple stellar systems which demands  
theoretical research on their origin and evolution, besides enabling
precise mass measurements.
With emission spanning the entire electromagnetic spectrum and some NSs showing
strange transient behaviour and even dramatic high-energy outbursts,
such incredible range and diversity was not only unpredicted, but in 
many ways astonishing given the perhaps naively simple nature of the
NS --- the last stellar bastion before the total collapse to a black hole.
But why do NSs exhibit so much `hair'? Even within the field there is
confusion about the sheer number of different NS class nomenclature.
For a recent review on the members of the NS zoo and the possible unification
of the various flavours, we refer to \citet{kas10} and references therein. An overview of
the formation and evolution of NSs in binaries is given, for example, in \citet{bv91,tv06}.

\begin{figure*}[b]
  \begin{center}
    \includegraphics[width=0.70\textwidth, angle=-90]{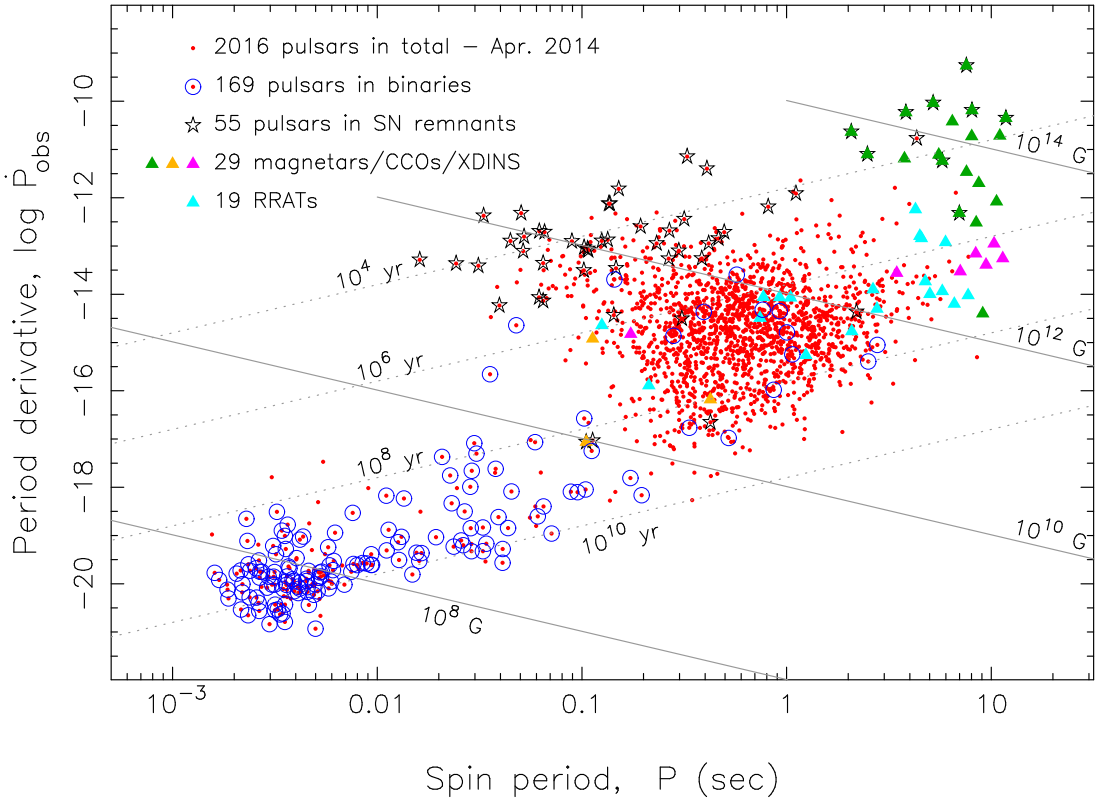}
    \caption{All currently known pulsars with measured $P$ and $\dot{P}$. Data from the 
             ATNF Pulsar Catalogue \citep{mhth05} -- April 2014.
             Lines of constant characteristic age and constant B-field are marked.}
  \label{fig:PPdot}
  \end{center}
\end{figure*}

In Figure~\ref{fig:PPdot} we have plotted all currently known radio pulsars
with measured values of spin period ($P$) and its time derivative ($\dot{P}$).
The differences in $P$ and $\dot{P}$ imply fundamentally different magnetic field strengths and ages for
each sub-group of the population. Treating the pulsar as a rotating magnetic dipole, one
may show \citep[e.g.][]{lk04} that the surface magnetic field strength
$B \propto (P \dot{P})^{1/2}$ and define a {\it characteristic age} $\tau_{\rm c} = P/(2\dot{P})$. 
The classic {\it radio pulsars} (red dots) are concentrated in the region with $P\simeq 0.2-2\;{\rm sec}$
and $\dot{P}\simeq 10^{-16}-10^{-13}$. They have magnetic fields of the order $B\simeq 10^{10}-10^{13}\;{\rm G}$
and a lifetime as a radio source of a few $10^7\;{\rm yr}$.
The population plotted in blue circles are binary pulsars and clearly indicate a
connection to the rapidly spinning {\it millisecond pulsars}.
The pulsars marked with stars indicate young pulsars observed inside, or near, their
gaseous supernova (SN) ejecta remnants. Then there are the ``drama queens'' of the
NS population: the {\it magnetars} which undergo high-energy bursts and are powered by their
huge magnetic energy reservoirs. Also plotted are the mysterious {\it rotating radio
transients} (RRATs), the {\it isolated X-ray dim NSs} (XDINS) and the {\it compact central objects} (CCOs).
In addition, we briefly discuss other exotic radio pulsars such as the
{\it intermittent pulsars}, the {\it black widows} and {\it redbacks}, and pulsars in triple systems.

A main challenge of the past decade was --- and continues to be --- to find a way to
unify this variety into a coherent physical picture. 
The NS zoo raises essential questions like: What determines whether a NS will be born with, 
for example, magnetar-like properties or as a Crab-like pulsar? What are the branching ratios for the various varieties, and,
given estimates of their lifetimes, how many of each are there in the Galaxy? Can individual NSs evolve from one species to another (and why)?
How does a NS interact with a companion star during and near the end of mass transfer (recycling)? And what
limits its final spin period?
Ultimately such questions are fundamental to understanding the fate of massive stars, close binary evolution
and the nature of core-collapse SNe, while simultaneously relating to a wider variety
of interesting fundamental physics and astrophysics questions ranging from the physics of matter in ultra-high magnetic fields,  
to the equation-of-state of ultra-dense matter and details of accretion processes. 

In this chapter we address the many faces of NSs with a strong emphasis
on radio pulsars and how the SKA will push this research area further to the forefront of
modern astrophysics. 
The structure of the chapter follows the NS population from its very youngest sources,
that still bear the imprint of their birth conditions, to the oldest recycled millisecond pulsars,
resulting from interactions with a companion star on a Gyr timescale. 
The bulk of the NSs are somewhere in the middle of their evolutionary journey, so that
ultimately detailed population synthesis studies of the full bounty offered by future SKA searches will
be key to a complete picture.
Whereas the chapter on ``A Cosmic Census of Radio Pulsars'' \citep{kbk+14} addresses more technical aspects and emphasizes what sources might be found, 
this chapter has more specific focus on the consequences of these discoveries and how they would relate to the general population of neutron stars
and our understanding thereof.

\section{Magnetars}\label{sec:magnetars}

Without doubt the most dramatic and radiatively unpredictable of NSs 
are the so-called ``magnetars,'' believed to be very young and the most highly magnetized
of the population. These objects, of which 26 are currently known
\citep{ok14}\footnote{See online magnetar catalogue at http://www.physics.mcgill.ca/~pulsar/magnetar/main.html},
have as their hallmark emission bright occasional super-Eddington bursts of X-rays and soft gamma rays.
The energy source for these bursts, as well as for the luminous X-ray pulsations observed from many in this class of objects, is believed
to be the decay of an intense internal magnetic field, which causes stresses and fractures of the stellar
crust \citep{td95,td96}. These fields are inferred from a variety of different arguments \citep[see,][]{td96} but arguably
most commonly from the spin period and period derivative measured from the X-ray pulsations via the conventional
dipole spin-down formula: $B=3.2 \times 10^{19} \sqrt{P \dot{P}}\;{\rm G}$, often yielding dipolar B-fields $>10^{14}$~G. 
Outstanding questions regarding magnetars include the origin of such high magnetic fields \citep{dt92}, what fraction of the
NS population is born as magnetars \citep{kk08}, and whether there is an evolutionary relationship between
magnetars and other NSs, such as high-magnetic-field radio pulsars (see Section~\ref{sec:high-B} below) or
isolated NSs \citep{tur09}.
Recently, \citet{khc+14} have suggested that magnetars may populate a group of high-mass X-ray binaries
observed in the Small Magellanic Cloud. 

Of the 26 known magnetars, only four have been radio-detected thus far \citep{crh+06,crhr07,lbb+10,efk+13}
in spite of sensitive searches of many others \citep{bri+06,chk07,lkc+12}.
The four solidly detected magnetars show interesting radio properties.
The first radio-detected magnetar, XTE J1810$-$197 \citep{crh+06}, 
became a bright radio pulsar only after its 2003 X-ray outburst, and showed a remarkably
flat radio spectrum, at some point 
being the brightest known radio pulsar above 20~GHz.
Extreme variability as well as a very high degree ($\sim\!100$\%) of linear polarization
were also observed \citep{ksj+07,crj+07} as well as dramatic pulse profile changes \citep{ccr+07}.
Meanwhile, the second-detected radio magnetar, 1E 1547.0$-$5408 \citep{crhr07}, has shown
properties quite similar to those of XTE J1810$-$197. 
A third radio-loud magnetar, PSR~J1622$-$4950, was discovered \citep{lbb+10} 
in apparent X-ray quiescence during a standard survey for radio pulsars,
hence making it the first radio-discovered magnetar.
It too was characterized by extreme variability, with epochs during which it was undetectable, and a flat or even inverted radio spectrum. 
Subsequent X-ray observations suggest that it is likely recovering
from an earlier X-ray outburst \citep{ags+12}, further suggesting that magnetar radio
emission is associated with post-outburst events. 

Very recently, a fourth radio magnetar, SGR J1745$-$29, was discovered via its X-ray emission, 
only 3$''$ from the Galactic Centre \citep{mgz+13}. The subsequent radio detection
\citep{efk+13,sj13} showed again high linear polarization, and the largest dispersion and rotation measures for any known pulsar.
In contrast to the other three radio-detected magnetars, this source appears to be far less radio variable, 
even as the X-ray emission fades \citep{kab+14,lk15}.
Interestingly, the high Faraday rotation as inferred from radio
observations argues for a dynamically significant magnetic field near the central
super-massive black hole, relevant to the physics of the accretion flow \citep{efk+13}.  
Importantly, a multi-wavelength pulse profile study \citep{sle+14} shows the
pulse is scatter-broadened by an order of magnitude less than was predicted by the
in-standard-use \citet{cl02} model, a conclusion supported by radio angular broadening
measurements \citep{bdd+14}.  
This is significant for potential SKA pulsar searches of the Galactic 
Center region \citep[see,][]{elc+14,kbk+14},
as it implies pulsars may be easier to find there than has been previously thought \citep{cl14}, particularly
at high radio frequencies such as SKA1-MID~band~5, even in spite of the very few detections thus far in relatively
high-frequency surveys \citep{bjl+11}. The radio detections of pulsars within the central
parsec of the Galaxy could allow unique dynamical tests that could constrain properties
of the central super-massive black hole as well as test theories of gravity \citep{pl04,lwk+12,elc+14}.

\subsection{The contribution of SKA}
The SKA will be of great use for the radio study of magnetars.
Current upper limits on radio observed, but yet undetected, sources are in the range of 0.01--0.04 mJy for
periodic emission and 0.1--10 mJy for single pulses, depending on pulse width \citep[e.g.][]{lkc+12}.
SKA sensitivities are up to an order of magnitude better than was previously possible, thereby greatly increasing the available
phase space for detection at a wide range of frequencies. The detection and/or improved upper limits
for additional sources
will help determine if the bulk are indeed radio quiet, as might be expected in some models
of pulsar emission \citep{bh01} or due to small beaming angles \citep{lkc+12}.
The SKA detection of radio emission
from a magnetar which previously appeared to have faded beyond observability in the radio band
could demonstrate persistent low-level radio emission.
A comparison of the radio properties of such a source in
radio quiescence with those when it was in outburst may be further constraining of models. 
Differences in polarization profiles between these two states could prove to be highly
effective diagnostics of geometric differences expected in twisted magnetosphere models \citep{tlk02,bt07}.
Regular monitoring of magnetars with SKA-type sensitivities could help determine whether their
hallmark bursts are present in the radio band as has been suggested in some burst models
\citep{lyu02}. 
Furthermore, some authors have suggested fundamental physics experiments with magnetar radio emission, 
such as axion detection using photon beam splitting \citep{gc11}, which may be facilitated with
more sources detected.

\section{High B-Field Radio Pulsars}\label{sec:high-B}

One question regarding the NS population is what distinguishes magnetars
from otherwise apparently conventional radio pulsars which appear, from their $P$ and
$\dot{P}$, to have magnetar-strength B-fields. Indeed currently
there are 9 known, otherwise conventional radio pulsars, having spin-inferred surface dipolar
magnetic field strengths above $4.4 \times 10^{13}$~G (a somewhat arbitrary limit derived
from the so-called ``quantum critical field'' $B = m^2 c^2 / \hbar e$), 
with the highest known being $B = 9.4 \times 10^{13}$~G for PSR~J1847$-$0130. 
Such fields clearly overlap, and in some cases exceed, those of 
magnetars, for which the smallest spin-inferred dipolar magnetic field is a
paltry $B=6.1 \times 10^{12}$~G \citep{rip+13}. Therefore, there is reason to expect
potential magnetar-like behavior in high-B radio pulsars. Indeed, PSR~J1846$-$0258 in the SN 
remnant Kes 73 has shown magnetar-like phenomenology \citep{ggg+08} 
in the form of a short-lived X-ray outburst simultaneous with a large rotation glitch. 
Deep X-ray observations of high-B radio pulsars
using telescopes like {\it Chandra} and {\it XMM-Newton} have also shown strong evidence for enhanced
thermal X-ray emission from high-B radio pulsars relative to that seen in lower-B radio pulsars
of comparable age \citep{km05,zkm+11,ozv+13}. This could be understood
in terms of magneto-thermal evolution in NSs, in which B-fields decay internally,
releasing energy and keeping the NS hot \citep{apm08,pvpr13}.

\subsection{The contribution of SKA}
The discovery of new high-B radio pulsars in surveys with the SKA has the potential to make important
progress in this field, primarily because of the paucity of high-B pulsars currently known. With new
examples, together with follow-up X-ray observations, particularly with future high-throughput X-ray
missions such as {\it ATHENA}\footnote{http://www.the-athena-x-ray-observatory.eu/},
additional data points can be added on NS
cooling curves \citep{ozv+13}. Given the expectations for the numbers of pulsars to be discovered with SKA,
and assuming the current ratio of high-B radio pulsars to the rest of the population ($\sim$0.4\%) continues to hold,
SKA should discover 50--100 new high-B sources for X-ray follow-up.

Moreover, regular monitoring of these high-B radio pulsars for timing purposes
is likely to reveal many spin glitches, particularly given how ubiquitous and dramatic such events are in magnetars
\citep{dkg08,dk14}:  spin-up and -down glitches have been observed and are generally accompanied by flux
enhancements as well as pulse profile changes and bursts. The timely detection of a glitch in a high-B
radio pulsar could yield an opportunity for rapid X-ray follow-up observations
to search for an X-ray outburst. 
Indeed, there already exist three examples of high-B radio pulsars exhibiting interesting pulse variability at glitch
epochs \citep{ggg+08,wje11,ksj13}. 
The discovery of new, young high-B radio pulsars could permit the measurement
of more braking indices.  The young radio pulsar PSR~J1734$-$3333 has $B=5\times 10^{13}$~G, and a braking
index of only $n=0.9 \pm 0.2$ \citep{elk+11}, the lowest yet measured for any radio pulsar and well below the canonical
$n=3$ expected for magnetic dipole radiation in a vacuum. Such a low braking index can be interpreted as
resulting from magnetic-field growth, suggesting that magnetars could be descendants of high-B radio pulsars.

\section{Central Compact Objects and Neutron Stars in Supernova Remnants}\label{sec:CCOs}

Another group of NSs are the central compact objects \citep[CCOs; see e.g.][]{psgz02}. 
These are isolated, non-variable point sources associated with supernova remnants (SNRs), seen in
thermal X-rays. So far they are without optical or radio counterparts
(although the latter are single-dish limits). CCOs have low X-ray
luminosities ($\sim 10^{33}\;{\rm erg}\,{\rm s}^{-1}$) and do not have associated pulsar wind nebulae, suggesting
that these stars are NSs which may not yet be active as pulsars. There
are currently eight confirmed CCO sources plus three candidates\footnote{Some authors also
  define 1E~161348$-$5055, the variable X-ray source associated with the RCW~103
  SNR as a CCO. However here we follow the definition of
  \citet{hg10} and take CCOs as having steady X-ray flux.} \citep{wsc+06,d08,gha13}
and multi-wavelength observing campaigns have been undertaken to
search for more sources in nearby SNRs~\citep{kgks06}. For three of the CCOs, the spin
characteristics ($P$ and $\dot{P}$) are known (see Figure~\ref{fig:PPdot}) from X-ray timing observations and indicate
surprisingly low magnetic fields ($\sim 10^{10}$~G, now in two cases, e.g. \citet{gha13}).

Despite the small number of known CCOs it is possible to estimate a minimum Galactic
birth rate of these sources using the ages estimated from the SNR expansion times. 
This yields $\beta_{\mathrm{CCO}}\sim 0.5 \;\mathrm{century}^{-1}$~\citep{gbs00}. CCOs are thus a significant
contributor to the so-called ``NS birthrate problem'' \citep{kk08}:
the problem that arises if one assumes all of the various NS
manifestations discussed in this chapter are evolutionarily
independent. Their cumulative birthrate does not tally with the 
estimated core-collapse SN rate for the Galaxy
and so the various groups must evolve into
one another along as yet uncertain evolutionary sequences. 
Moreover, the spin-inferred magnetic fields of CCOs are so small that
their expected lifetimes above the death-line in the $P$-$\dot{P}$~diagram
are tens to hundreds of millions of years; this conflicts strongly with the apparent
paucity of sources in this region of the $P$-$\dot{P}$~diagram, {\it if} the typical CCO is a radio emitter.
CCOs are thus a very important part of the puzzle on NS evolutionary pathways. 

\subsection{The contribution of SKA}
A deep investigation with both SKA1 and SKA2
could determine once and for all whether CCOs are radio emitters. 
In addition to the CCOs, there are currently 55~pulsars with SNR associations. 
Figure~\ref{fig:PPdot} shows that these
sources are predominantly ``young'' with $\tau_{\mathrm{c}}=10^{4-5}$~yr,
consistent with the short lifetimes of visible SNRs. The
``older'' sources with SNR associations are either chance superpositions \citep{bnk14}
or are in fact young, i.e. having been born with a slow birth period causing 
their characteristic age to be much larger than their true age. The SKA will enable unprecedented deep searches
of all of the Galactic SNRs visible in the SKA-sky and will firmly
detect all of those showing even minute levels of radio pulsar emission. 
Timing these pulsars with the SKA will allow measurement of their spin
characteristics and thence enable us to ``rewind'' the spin history of
these stars to determine a more realistic birth period distribution for
pulsars, a notoriously difficult aspect of pulsar evolution to model \citep{fk06,nsk+13}.

\section{Rotating Radio Transients}\label{sec:RRATs}

Rotating radio transients (RRATs) are a class of pulsars from which pulses are only sporadically detectable \citep{mll+06}. 
The appropriate definition of this class of objects, and indeed whether a definition is even necessary, has been debated \citep{km11}. 
Here, we define RRATs as NSs which were originally detectable only through their single pulses 
and not through a periodicity search (though we exclude the first pulsars discovered before the regular application of periodicity searches). 
Using this definition, there are roughly 100 known RRATs\footnote{See the online catalogue of RRATs at http://astro.phys.wvu.edu/rratalog}. 
Their periods range from 0.125~s to 7.7~s, with a mean of 2.6~s. A fraction ($\sim\!20$\%) of the RRATs do not yet have determined periods 
due to a small number of detected pulses. 
These periods show some evidence of being longer than those of other pulsars.
It is unclear if longer-period pulsars become sporadic emitters, or if this is a 
selection effect due to the smaller number of pulses in an observation --- and hence, for longer period pulsars, 
higher likelihood of being more detectable in a single-pulse search.

Measuring period derivatives for RRATs often requires much more observing time than for normal pulsars due to their sporadic emission. 
Therefore, only $\sim\!25$\% of the RRATs have measured period derivatives (cf. Figure~\ref{fig:PPdot}). These do not appear to be significantly different than the 
period derivatives of other pulsars, but due to the longer periods, the inferred surface dipole magnetic fields of the RRATs are possibly higher than 
for other pulsars, with a mean of $3\times 10^{12}\;{\rm G}$. 

When the discovery of the RRATs was first announced \citep{mll+06}, it was believed that they were a new class of radio-emitting NSs. 
Under this assumption, their sporadic properties imply that for each detected RRAT, there are many more undetected objects, 
leading one to infer a very large population of these sources, and hence NSs, in the Galaxy \citep{mll+06}.
This issue was further explored and \citet{kle+10} found that, assuming that the RRATs form a population distinct  
to other NSs, the total inferred number of Galactic NSs is discrepant with Galactic SN rate estimates. 
One solution to this problem is if RRATs are a part of the same population as other pulsars.

This assumption seems reasonable, given the similarity between the properties of RRATs and other pulsars, and the apparent continuum in 
intermittency bridging these two (perhaps not distinct) populations. Studies of the single pulses of RRATs indicate that their brightness 
temperatures are no greater than the single-pulse brightness temperatures for other pulsars. In addition, the pulse shapes and widths of RRAT pulses 
are similar to those of other long-period pulsars. Furthermore, some RRATs were initially detectable only through their single pulses, 
but with observations at a lower frequency or with a more sensitive telescope, they are detectable through periodicity searches like normal pulsars. 
It is therefore plausible to assume that RRATs are simply pulsars which sit on the tail end of a continuous intermittency spectrum 
and are not part of any special population.

\section{Intermittent Pulsars}\label{sec:IMpsrs}

\begin{figure}
  \begin{center}
    \includegraphics[width=0.55\textwidth, angle=0]{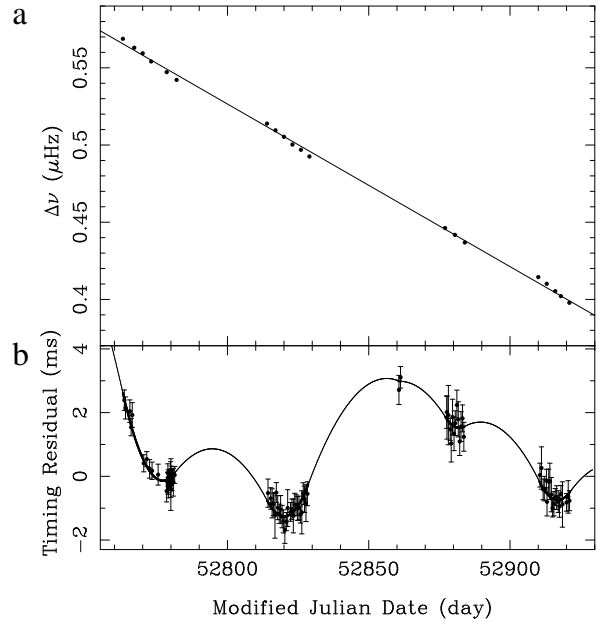}
\end{center}
\caption{\label{fig:1931} Changes in the spin-rate of PSR~B1931+24
  (upper panel) when `on' (radio bright). The rotational frequency
  is clearly decreasing faster (i.e. the gradient is steeper) in the
  `on'-state compared to the average spin-down rate (indicated by
  the solid line), implying that the pulsar is spinning down slower in
  the `off' (radio quiet) state. Taking the switch between two
  different slow-down rates into account, the apparent noisy timing
  behaviour can be explained perfectly (lower panel). Figure from
  \citet{klo+06}.}
\end{figure}

While there has been progress in the theoretical understanding of the
magnetospheric processes in pulsars \citep[e.g.][]{ckf99,spi06}, some of the
most important experimental insight comes from the study of a class of
NSs called {\it intermittent pulsars}. The archetype of this
relatively recently discovered class is radio pulsar B1931+24
\citep{klo+06,ysl+13}, which is only active for a few days between
periods of roughly a month during which the pulsar is not
detectable. The fact that pulsars are able to switch on and off is
something that has been known for a long time and is often observed at
much shorter timescales, in which case it is known as {\it nulling}
\citep[e.g.][]{bac70,wmj07}. The exciting result for the intermittent
pulsars is that these sudden switches are accompanied by changes in
the spin-down rate of the NS rotation, suggestive of
significant changes in the torque generated by magnetospheric
currents. For PSR~B1931+24 the spin-down was shown to be 50\% larger
when the radio emission of the pulsar is in the {\it on-state} compared to
the {\it off-state} \citep{klo+06}, explaining at the same time the
seemingly noisy timing behaviour for this source (see
Figure~\ref{fig:1931}).

Besides PSR~B1931+24, there are only two other
intermittent pulsars known which change their spin-down rate. 
For PSR~J1832+0029 \citep{llm+12} the spin-down rate changes by a factor of
1.8, while for PSR~J1841--0500 \citep{crc+12} the spin-down rate
changes by a factor of 2.5. All three intermittent pulsars appear to be part of the
bulk of the pulsars in the $P$-$\dot{P}$~diagram.
At first, it may seem somewhat surprising that radio emission is
linked to the spin-down rate of pulsars. The amount of energy in the
radio emission is tiny in comparison to the total loss rate of the
rotational energy of the NS (i.e. $\dot{E}$), so one could expect the
radio emission to be an indicator of higher order effects rather than
fundamental processes within the pulsar
magnetosphere. Nevertheless, the radio emission appears to be a tracer
of powerful currents in the magnetosphere.

The behavior of intermittent pulsars can be explained by a relatively
simple two-state magnetospheric model \citep{klo+06,lst12}. In one of
the states the open field line region is filled with a co-rotating
plasma which as a consequence should have the Goldreich--Julian
density \citep{gj69}. This magnetospheric state can in good
approximation be described by a force-free magnetospheric solution
\citep{spi06}. This plasma is generated by the same mechanism which
produces the radio emission observed during the on-state. When this
mechanism switches off both the radio emission and plasma production
ceases, resulting in the open field line region to get depleted from
plasma. This vacuum state will have a different spin-down rate by a
factor which depends on the magnetic inclination angle \citep{lst12},
which can explain the range (1.8--2.5) in observed factors by which the spin-down
rate is changing.

An alternative model has been suggested \citep{tim10} in which 
the switching between stable magnetospheric states causes 
different geometries or/and different distributions of currents
which can explain the persistent ``nulling'' during the off-states.
Furthermore, it has become clear that there are more subtle ways in which the radio
emission and spin-down rate are linked. There are a number of objects
which show pulse profile changes concurrent with changes in the
spin-down rate \citep{lhk+10}. This has led to the suggestion that all
the so-called timing-noise (deviations from a simple spin-down model)
could be ascribed to switches between magnetospheric states. These
more subtle changes in the emission result in more subtle changes in
spin-down, which can be less than $1\%$. One should note, however,
that most of these state changes (e.g. the so-called ``moding'')
happens on time-scales of minutes to hours and are therefore too fast
to measure (and, hence, separate) correlated changes in spin-down rate \citep{ysw+12}. 

\subsection{The contribution of SKA}
The SKA can help in several ways to make progress in our understanding
of the pulsar magnetosphere using intermittent pulsars \citep[see also][]{kja+14}. 
Firstly, we will progress by discovering more intermittent pulsars. 
This will lead to additional measurements of the changes in the spin-down rate between the on-
and the off-state for many more sources, thus understanding better the magnetospheric state changes.
For this it will be important,
however, that repeated passes are performed over given search
areas. The time between these passes should vary from hours and days
to weeks and months, similar to the range of time scales observed for
both intermittent and nulling pulsars (as well as RRATs, see Section~\ref{sec:RRATs}). 
Indeed, an important question which the SKA will help to address is what the difference (if any) is between
intermittent pulsars and nulling pulsars, 
as the main difference appears to be the duration of the off-state. 

It is important to note that the large sensitivity of the SKA
will not necessarily mean that we can detect much smaller changes in
the spin-down rate or changes which are shorter in duration. 
However, the high sensitivity of the SKA can put better (important)
limits on the presence of radio emission during the off-state. It is
easy to confuse nulls with very weak emission, as has been shown for
instance for PSR~B0826--34 \citep{elg+05}. The presence of very weak
emission during the off-state of intermittent pulsars would show
that there are at least some currents present in the pulsar
magnetosphere at all times. 
Additionally, the multi-beaming capability of the SKA will allow
a dense monitoring of the known intermittent (and other) pulsars.
The much larger possible cadence will allow us to determine the 
exact switch times between the states more accurately.
This is important for two reasons: i) we can probe the
plasma conditions much better by constraining
the currents present in the magnetospheres
and correlate the results to other properties of pulsars, and
ii) we can potentially use the information to correct for timing noise in non-recycled pulsars,
hence, opening up the possibility to use them also for 
high precision experiments such as gravitational wave detection \citep{lhk+10}.

\section{Isolated Pulsar Evolution on the $P$-$\dot{P}$~Diagram}\label{sec:PPdot}

A common approach to understand the distribution of pulsars in the $P$-$\dot{P}$~diagram (Figure~\ref{fig:PPdot}) 
is to take a prescription for the spin evolution
of NSs (often a simple magnetic dipole model) and 
evolve pulsars forward from some birth distributions
in $P$ and $B$ in time to find the present-day distribution. Ideally,
after accounting for observational selection effects \citep{ec89}, this distribution would be statistically
indistinguishable from the one that is observed. The major challenge
facing all these studies is to decouple various competing and covariant
effects that shape this distribution. 
Assuming that radiation beaming \citep{tm98} and the selection effect
modelling process \citep{fk06} are reasonably well understood, 
the main factors involved are discussed in turn below.
A less model dependent approach is to
compute the flux or ``current'' of pulsars through the $P$-$\dot{P}$~diagram
by taking into account detection biases \citep{pb81,vn81,nar87}.
Recently, \citet{lgy+12} have attempted to categorize
sub-populations within the distribution using Gaussian mixture models.
The main factors for shaping the pulsar distribution are:

\begin{itemize}
\item{{\bf Neutron star birth parameters.} While the overall consensus
  on the distribution of B-fields seems to favour a
  single-component log-normal distribution, much debate has taken
  place over the form of the period distribution.  The view put
  forward by early studies \citep[e.g.][]{lmt85} which favoured a model in
  which all pulsars were born spinning rapidly was challenged by the
  idea of an ``injection'' of pulsars into the population with birth
  periods of around 0.5~s \citep{vn81,nar87,no90}. Current studies
  \citep{fk06,rl10} favour a broad normal distribution with a mean of
  0.3~s and standard deviation of 0.5~s. 
  Independent constraints on the initial period
  distribution from pulsar--SN remnant pairs \citep{pt12}
  suggest a mean and standard deviation of 0.1~s. 
}

\item{{\bf Magnetic field strength evolution.} Solving the differential
  equation $B \propto (P \dot{P})^{1/2}$ to obtain the evolution of $P$ and
  $\dot{P}$ with time in the magnetic dipole model is most readily
  achieved assuming that $B$ is constant. Such models provide good
  matches to the $P$-$\dot{P}$ distribution \citep{bwhv92,fk06}. However,
  models which invoke a very modestly decaying magnetic field \citep{ppm+10} 
  are also consistent with the data. Earlier theoretical work \citep{tk01} also
  reached similar conclusions.
}

\item{{\bf Magnetic inclination evolution.} Within the frame of
  a rotating magnetic dipole in a vacuum, the torque acting on the
  spinning NS is proportional to $B^2\,\sin ^2 \alpha$, where $\alpha$
  is the angle between the spin and magnetic axes. This  
  relationship implies that magnetic field decay mentioned in the
  previous item can also be mimicked by a decay in the inclination
  angle, $\alpha$. Several independent studies  \citep{tm98,wj08,ycbb10}
  conclude that $\alpha$ decays with time.
  An outstanding issue \citep{rl10} is how this decay translates to a change in the
  braking torque on the pulsar. While \citet{tk01} suggest that a
  decay in the braking torque is required by the $P$-$\dot{P}$
  distribution, as \citet{rl10} demonstrate, Monte Carlo simulations
  assuming magnetic alignment cannot satisfactorily explain the
  observed sample.
}

\item{{\bf Non-standard magnetic dipole evolution.} One possible solution to the
  alignment issue discussed above is to evolve the pulse period
  according to a different spin-down law. A revised model has been proposed \citep{cs06} 
  with different evolutionary tracks which allow for a
  constant magnetic field and alignment. Detailed Monte Carlo
  models \citep{rl10} have, however, so far not been able to reproduce
  the $P$-$\dot{P}$~diagram using this framework.
}

\item{{\bf Luminosity ``law'' and evolution.} In the absence of a well
  developed theoretical framework for understanding the radio emission
  of pulsars, the luminosity $L$ is often calculated from a power-law
  expression of the form $L \propto P^{\alpha} \dot{P}^{\beta}$. This
  approach has been adopted by numerous 
  studies \citep[e.g.][]{lmt85,sto87,no90,fk06}, with quite different values for the
  exponents $\alpha$ and $\beta$. It is noteworthy \citep{lbdh93}
  that the adopted values for $\alpha$
  and $\beta$ significantly impact the conclusions regarding
  population parameters. For example, combinations which favour high
  luminosity with short periods generally support injection into the
  population \citep{nar87}. Recent work by \citet{blrs14} find $\alpha
  \simeq -1.4$ and $\beta \simeq 0.5$ which is very similar to the
  relationship found for the gamma-ray pulsar population
  \citep{pmc+13}. This combination of parameters leads inexorably to
  the conclusion that the radio luminosity decays significantly with
  time. Expressions which describe these trends are provided by \citet{blrs14}
  and make predictions for large-scale surveys, such as the SKA. 
}

\item{{\bf Death-lines and valleys.} The dearth of pulsars in the lower
  right-hand side of the $P$-$\dot{P}$~diagram has long been associated
  with a physical limit on radio emission; for a review see \citet{cr93}.
  As such, death-lines (or valleys) are frequently included in pulsar
  population syntheses. Very recently, a reasonable reproduction of
  the $P$-$\dot{P}$~diagram has been carried out without the adoption of
  a death-line, or any luminosity law \citep{szm+14}. Instead, the study invoked
  an inverse correlation between radio efficiency and age. This approach 
  appears to argue against any dependence of radio luminosity with $P$ 
  and $\dot{P}$.
}

\item{{\bf Braking indices.} Additional 
measurements of the braking indices of pulsars \citep{lkg+07} would help to
pin down the directionality in the flow of pulsars in the diagram.
Indeed, some of the braking indices measured so far for young pulsars
are not consistent with the locations of older pulsars and hint
at the idea that some radio pulsars may evolve into magnetars \citep{elk+11}. 
As noted by earlier authors \citep[e.g.][]{tk01}, 
a change in the braking index is required throughout the lifetime of a pulsar.
This implication has yet to be fully addressed by population studies.
}
\end{itemize}

It is evident from the preceding discussion that many uncertainties
remain in our understanding of the $P$-$\dot{P}$~diagram. The larger 
sample of pulsars provided by SKA surveys 
will undoubtedly help to clarify many of these issues, including  
the shape and the faint-end of the pulsar luminosity function. While many authors \citep[e.g.][]{lmt85,lbdh93}
adopt power-law models of pulsars luminosities, recent work \citep{fk06,rl10,blrs14}
suggests that the parent population is more log-normal in nature.

\section{Millisecond and Binary Pulsar Surveys with SKA}\label{sec:MSPs}

\begin{figure*}[t]
    \includegraphics[width=0.60\textwidth]{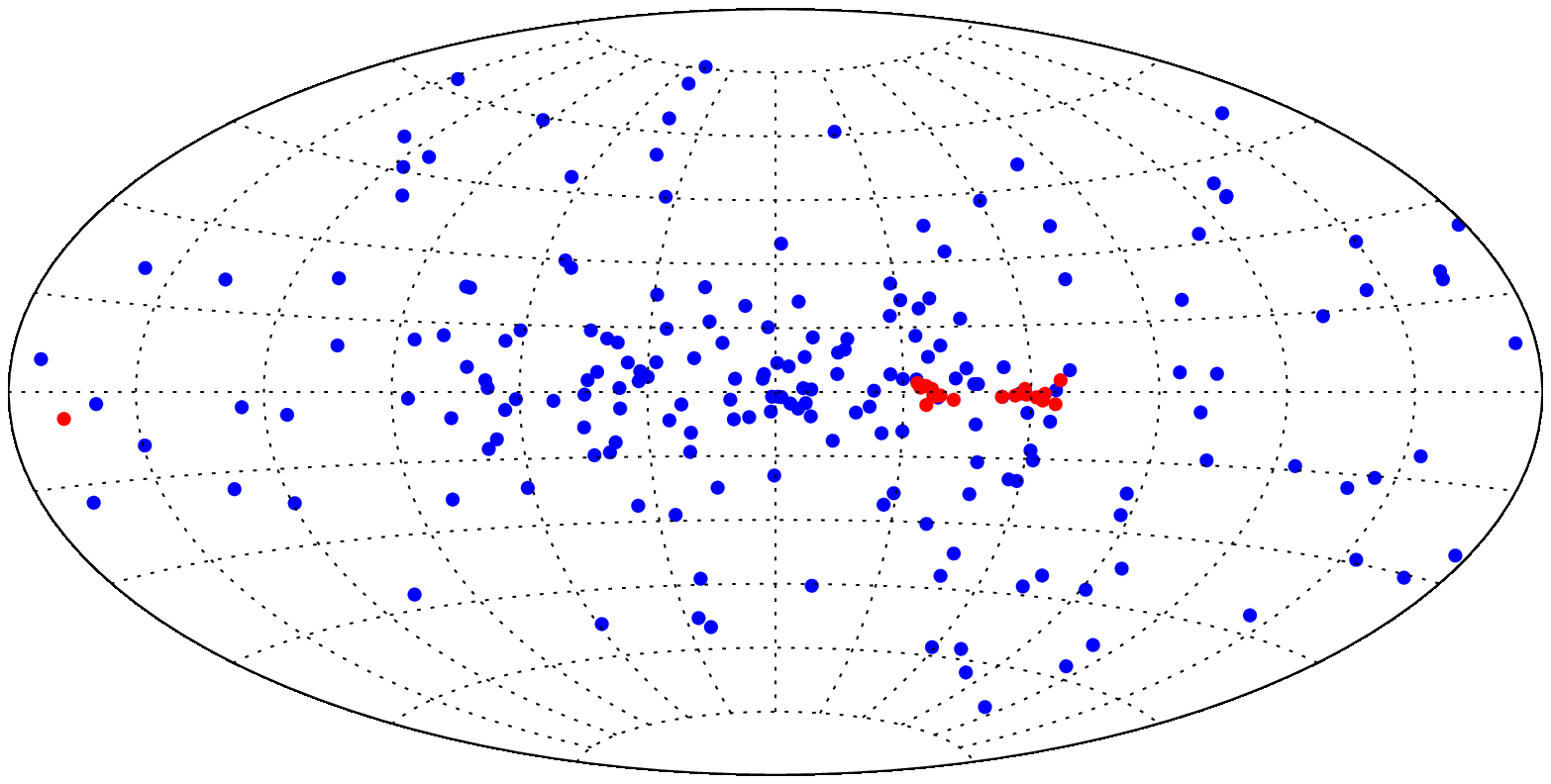}\includegraphics[width=0.38\textwidth]{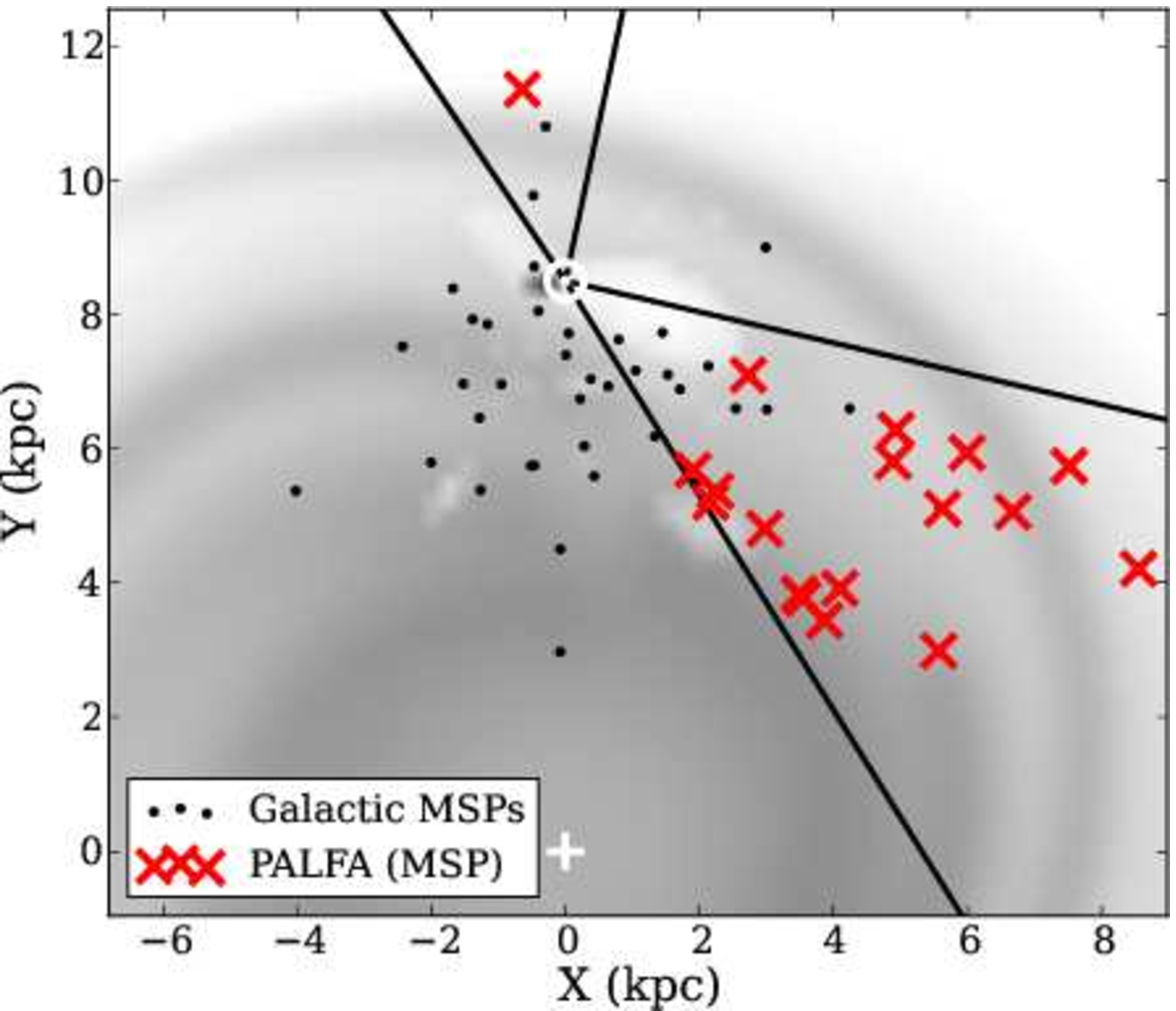}
  \caption{Left: Locations of the 216 known MSPs outside of globular clusters. 
  Discoveries from the PALFA survey at Arecibo are shown in red. 
  Right: Top down view of the galaxy with MSP locations 
  marked. Again, red symbols denote PALFA MSPs \citep[reproduced from][]{laz13}, 
  which are systematically more distant than the previously known population. 
  The solid lines show the portion of the plane visible from Arecibo. 
  SKA1 will expand that to cover nearly the entire Galaxy.}
  \label{fig:galmsps}
\end{figure*}

The large total collecting area, wide range of frequencies, and rapid
survey speed available with SKA1 will revolutionize our
understanding of the MSP population in the Milky Way.
Currently, there are $\sim\!300$ MSPs (defined here as $P<30$ ms) known in
the Galaxy, including those associated with globular clusters \citep[see the chapter by][]{hpb+14}. 
Their distribution on the sky and in the Galaxy is shown in Figure~\ref{fig:galmsps}.
The primary SKA project that will drive our understanding of the MSP
population is the SKA1-MID pulsar survey, which will cover 36,000 
sq.\,deg. of sky. This is only possible with the large collecting area and
very fast survey speed of SKA1-MID. 
Surveys at high latitude may reveal new bright, stable pulsars
that will contribute to pulsar timing array projects, and will fill out our understanding
of the low end of the MSP luminosity function ($<0.1\;{\rm mJy\,kpc}^2$ at 1400~MHz).
In the plane SKA will explore a larger volume of the Galaxy than ever
before to find rare systems and constrain the Galactic population of MSPs (see Section~\ref{sec:binaries}).
The SKA1-MID pulsar survey, as defined in the SKA1 System Baseline
Design is of similar sensitivity to the ongoing PALFA survey being done
with the Arecibo Telescope, with flux limits in the neighborhood of 
$40\;\mu$Jy. The big difference is that the SKA1-MID survey extends that
sensitivity over the full visible sky, while PALFA is  
constrained to narrow strips of the Galactic plane visible from
Arecibo (see Figure~\ref{fig:galmsps}).

These very deep surveys are sensitive to pulsars at large
dispersion measure, because of the relatively high observing
frequencies, and thus probe a much larger volume of the Galaxy than previous
large-area surveys. For directions in the
plane, the volume sampled increases as the square of the distance
probed. Figure~\ref{fig:galmsps} shows that PALFA is seeing a much
more distant population than other surveys. SKA1-MID will expand this
over nearly all of the plane, particularly towards the Galactic Bulge,
where large numbers of low-mass X-ray binaries (the progenitor systems
for MSPs) are known to reside.  This will help
constrain the total number of MSPs in the Galaxy and
their density as a function of galactocentric radius, which is
important for population modeling. 

From an evolutionary point of view, many of the most interesting pulsars are found in tight
relativistic orbits, allowing the measurement of post-Keplerian parameters. 
This implies that the pulsar will be accelerated for most of its orbit,
requiring either short integration time with high sensitivity or
sophisticated acceleration processing. 
Compensating for a sensitivity loss of about 10\% (with respect to the baseline design)
with increased observing time results in a doubling
of the processing requirements to find the same relativistic binary
pulsars. Hence, conducting pulsar searches with SKA1-LOW and its
exceptional collecting area is a promising method for finding 
binary pulsars away from the Galactic plane. Indeed, our
simulations show that a combined SKA1-MID and SKA1-LOW survey
targeting regions on and off the Galactic plane respectively finds
approximately 20\% more pulsars than can be found with SKA1-MID alone \citep{ks14,kbk+14,ks14}.
Binary pulsar masses (Section~\ref{sec:NSmass}) can only be determined if regular
timing observations can be done. Though
some sources may be bright enough to follow-up with other radio
telescopes, it is likely that most sources will require the
SKA. Sub-arraying can help alleviate the observing load for sources
that are easily detectable with a fraction of the collecting area. It
may also be advantageous in some cases to run pulsar timing
experiments in parallel with imaging observations to boost the
observing efficiency.

When SKA2 is realized, the additional sensitivity will enable considerably more breakthrough science in this field. 
SKA will not only provide further sources to characterize the MSP and binary pulsar population 
(in particular via a high frequency survey in the Galactic bulge and central region), also the detection of giant pulses in
extra-galactic MSPs (e.g. in M31 and IC~10) may be realistic.

\section{Pulsar Recycling and Formation of Exotic Binaries and Triples}\label{sec:binaries}

Radio pulsars have been discovered in binary systems with a variety of companions: white dwarfs (WDs), NSs, main sequence stars, and even planets.  
The vast majority of these systems contain an MSP with a helium WD companion. 
Based on stellar evolution theory, it is expected that pulsars
could also be found with a helium star or a black hole companion. These systems have not yet been discovered
but it is likely that the SKA will reveal such pulsars.
There is a growing number of systems
where the companion star is being ablated by the pulsar wind, the so-called {\it black widows} and {\it redbacks} \citep{rob13}. 
This is evidenced by the radio signal from the pulsar being eclipsed for some fraction of the orbit \citep{fst88,sbl+96,asr+09}.
These companions are low-mass semi-degenerate stars with a mass between $0.02-0.3\;M_{\odot}$ \citep{rob13,bvr+13}.
Whereas some studies clearly indicate that the two populations of black widows and redbacks are distinct \citep{ccth13}
others argue in favour of a continuous evolutionary link \citep{bdh14}. 
This question is important to settle; 
as well as the long standing issue about whether they are the progenitors of the isolated MSPs \citep{rst89,pos13,db03}.  

In recent years, a few binary pulsars with peculiar properties have been discovered and which 
indicate a hierarchical triple system origin --- e.g. PSR~J1903+0327 \citep{crl+08,fbw+11,pvvn11}.
In 2013, two puzzling MSPs were discovered in eccentric binaries: PSR~J2234+06 \citep{dsmb+13} and PSR~J1946+3417 \citep{bck+13}.
These two systems might also have a triple origin. However, their eccentricities and orbital periods have led to the suggestion of an
alternative hypothesis of direct MSP formation via a rotationally delayed accretion-induced collapse of
a massive WD \citep{ft14}.
Whereas the origin of these systems remains unclear, a triple system MSP with two WD companions (PSR~J0337+1715)
was announced earlier this year by \citet{rsa+14}.
This amazing system must have survived at least 3~phases of mass transfer and one SN explosion and challenges current
knowledge of multiple stellar system evolution \citep{tv14}. 
To understand the underlying physics it is therefore important to find more of these exotic systems. 

Surveys with SKA1-MID (including a coverage of the Magellanic Clouds) will indeed find many more rare systems. 
It is likely that SKA1 will increase the population of known pulsars and MSPs by at least a factor of 5 and thus greatly
increase the number of systems which can be used to study the evolution of pulsars that are in, or have been members
of, multi-stellar systems. 

MSPs obtain their rapid spin (and weak B-field) via a long phase of accretion of matter from a companion star 
in a low-mass X-ray binary \citep[LMXB;][]{bkh+82,acrs82,rs82,bv91,tlk12}. 
This is evidenced by the high incidence of binaries found among these fast spinning pulsars 
(see Figure~\ref{fig:PPdot}). 
This model was beautifully confirmed by the discovery of the 
first millisecond {X}-ray pulsar in the LMXB system SAX~1808.4--3658 \citep{wv98}, 
and more recently by the detection of the so-called transitional MSPs
which undergo changes between accretion and rotational powered states \citep{asr+09,pfb+13}. 
Many details of the recycling process, however, remain unclear. Some of the most
important issues \citep[e.g.][and references therein]{tlk12} are:
the accretion-induced decay of the surface magnetic field; 
the maximum possible spin rate of an MSP;  
the Roche-lobe decoupling phase; 
the accretion torque reversals and the spin-up line; and 
the progenitors of the isolated MSPs. 

\begin{figure*}[t]
  \begin{center}
    \includegraphics[width=0.46\textwidth, angle=0]{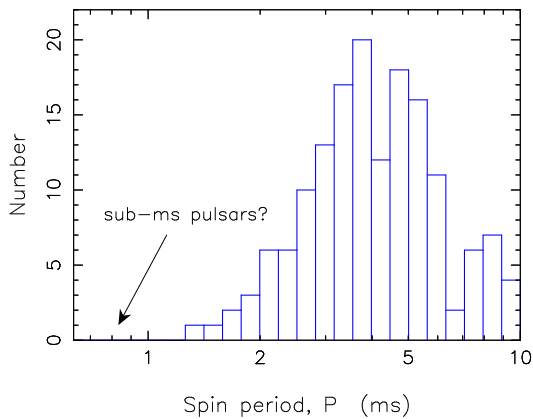}
    \caption{      The observed spin period distribution of MSPs. Data taken from the
                   ATNF Pulsar Catalogue \citep{mhth05} -- version 1.49, April 2014.
                   The discovery of sub-ms MSPs with the SKA would have a huge scientific impact. }
  \label{fig:Pspin}
  \end{center}
\end{figure*}

There is empirical evidence that the surface magnetic field strength of NSs
decays as a consequence of accretion, but the exact reason for this process is not well understood 
\citep{bha02}. Nor is it known what dictates the fastest possible spin rate of a radio MSP. 
In Figure~\ref{fig:Pspin} we have shown the spin period distribution of radio MSPs.   
Does the equation-of-state (EoS) for dense nuclear matter allow for the existence of sub-ms pulsars?
Or is the current spin frequency limit, slightly above 700~Hz \citep{hrs+06},  
set by the onset of gravitational wave (GW) emission during accretion \citep{cmm+03}, 
or subsequent r-mode instabilities? Alternatively, the spin-up torque might saturate 
due to magnetosphere--disc conditions, thus preventing sub-ms MSPs to form. In addition,
it has been demonstrated \citep{tau12} that during the final stage of mass transfer,  
MSPs may lose up to 50\% of their rotational energy 
when the donor star decouples from its Roche~lobe. 

\subsection{The contribution of SKA}
As described in detail elsewhere in this book, the large scale surveys performed with SKA1 will lead to 
a significant increase in the number of catalogued MSPs. That will also include several new exotic (and
therefore intrinsically rare, possibly unprecedented) objects, 
suitable to directly address some of the key-questions reported above: i.e. the discovery of even a single MSP
spinning at a period well below 1~ms would provide strong constraints on the NS structure and 
the EoS of dense nuclear matter. In the context of evolutionary studies, it will be particularly interesting 
to use SKA1-MID for performing a deep search for pulsars at a higher frequency (between 2 and 3~GHz) 
than ever for an all-sky survey. In fact, at those frequencies, the absorbing/eclipsing effects of the plasma 
released by the companion star are significantly reduced  with respect to what happens at the lower 
frequencies typically used in all the past large scale surveys. That could finally unveil the radio signal 
from several ultra close-orbit binary pulsars in the last phases of their recycling, i.e. the transitional MSPs
which are among the most precious systems for investigating the still unanswered issues discussed in this section. 

SKA2 will then provide an extremely large population of recycled pulsars, eventually enabling a statistically 
sound comparison between the properties of the various classes of binary and isolated MSPs,  
thus firmly establishing relationships and evolutionary connections. Similarly, it has been 
shown in several works \citep[e.g.][and references therein]{pcg+99,tlk12} that the availability of a large
population of MSPs (maybe including still unknown MSPs with surface B-fields below $10^7\;{\rm G},$ or MSPs
with relatively high B-fields) would constrain the physical processes during the last stages of accretion.  
This includes new knowledge on the evolutionary phases experienced by a pulsar close to the spin-up line and/or the occurrence of 
GW emission from the accreting NS, as well as the radio emission mechanism in old NSs (i.e. the death-line for low B-field pulsars).

\section{Double Neutron Star and Neutron Star/Black Hole Systems}

In 1974 a new class of radio pulsars was discovered with the detection of the first double NS system \citep{ht75}. 
Today we know a dozen double NS systems. 
The formation and evolution of such relativistic binaries (tight systems with massive compact stars) is a
key topic in modern astrophysics with many applications to fundamental physics in general.
In particular, double NS systems are important for:
pre-SN binary evolution and explosion physics, measurements of NS masses, testing theories of gravity, and 
their mergers are prime candidates for detection of high-frequency gravitational waves with LIGO/VIRGO within the next few years. 

Double NS binaries are important for both stellar astrophysics and SN explosion modelling since their observed orbital 
characteristics represent a fossil relic of the last (second) SN explosion. Hence, from observations of double NS binaries 
we can place constraints on both the pre-SN conditions and the explosion mechanism itself.
Many of the systems have rather small eccentricities and small masses of the
last (second) formed NS. This is indeed expected if these SNe were either caused by electron capture SNe \citep{plp+04} 
and/or a result of pre-SN ultra-stripping of the collapsing star \citep{tlm+13}. 
However, also higher birth masses of NSs are expected from stellar evolution \citep{tlk11} and the many new double NS binaries
anticipated to be discovered by the SKA will finally probe the full distribution of NS birth masses in these systems
(see Section~\ref{sec:NSmass}). The SKA is very likely to discover more double pulsar systems like
PSR~J0737$-$3039A/B. From observations \citep{bkk+08,bkm+12} and modelling \citep{lt05} of
interactions, via eclipse light curves, between the emission beam of one pulsar with the magnetosphere
of its companion pulsar, detailed knowledge can be obtained with respect to the
magnetospheric structure \citep[see also][]{kja+14}, the plasma properties, and relativistic precession of pulsars.

With the many new pulsar discoveries to be expected with the SKA there is realistic hope that they will also
include a few long-sought-after NS--black hole (NS--BH) binaries. NS--BH binaries have the potential to revolutionize 
gravity tests \citep{ksm+06,wex14,ssa+14} and will also bring new insight to the final stages of massive stellar evolution and SNe.  

Understanding the NS--NS and NS--BH populations is important since their merger events are prime candidate
sources for the upcoming detection of high-frequency gravitational waves with LIGO/VIRGO.  
The event rate of detections is expected to be of the order 0.1--10 per week \citep{vt03,aaa+10} and this rate
should match the observed distributions of orbital separations and eccentricities among Galactic
NS--NS and NS--BH sources. 
The LIGO/VIRGO detection rate is based on the Galactic merger rate of NS--NS and NS--BH systems which is of the order a~few$\;{\rm Myr}^{-1}$. 
However, this number is highly uncertain due to the many complex factors involved in computing the formation rate of these systems \citep[e.g.][]{vt03}.

\section{Neutron Star Masses}\label{sec:NSmass}

The masses of NSs play a vital role in a number of aspects pertaining to their structure, formation and evolution. 
So far, all pulsar mass measurements rely on one form or another of dynamical proxies offered by binary systems. 
The starting point is pulsar timing, which provides a very accurate determination of the mass function of the system and links 
the orbital inclination to the two masses. Supplementing it with two additional constraints therefore breaks the degeneracy. 
This can be achieved in various ways: i) Relativistic binaries --- usually double NS systems --- allow for the measurement of post-Keplerian parameters 
(such as the Shapiro delay and the periastron advance) that each yield an extra constraint, see \citet{ssa+14}; 
ii) Binaries containing an optically bright companion may provide its projected radial velocity via spectroscopy, 
a companion mass determination \citep[e.g. via the atmosphere modelling of WDs,][]{vbjj05}, 
or a measure of the mass ratio and orbital inclination via light curve modelling \citep[e.g.][]{bvr+13}.

Since any given EoS \citep[see e.g. the chapter by][]{wxe+14} sets a maximum mass stability limit, hunting for massive pulsars can provide an efficient way of constraining its nature. 
With a significant sample of NS masses, however, one can 
statistically constrain their EoS. Current observational evidence \citep{dpr+10,afw+13} disfavours a ``soft'' EoS, which cannot reach much 
beyond the measured $2\,M_\odot$.

Pulsar masses are also fundamental tracers of their formation and subsequent evolution via accretion. 
The birth mass of a NS relates to its formation and reflects the SN physics \citep{tlk11,fsk+13,fsk+14}.
Hence the presence of multiple modes in the birth mass distribution will probe their formation channels (i.e. iron core-collapse
SNe and electron capture SNe).
The mass distribution of NS--NS and NS--WD binaries was recently analysed \citep{kkdt13} and it was concluded that they peak at 
$1.33\,M_{\odot}$ and $1.55\,M_\odot$, respectively, with the first having a nearly 
symmetrical spread of $0.12\,M_\odot$ (1-$\sigma$) while the other one being skewed towards higher masses.
This arises from the fact that the observed pulsar masses are convoluted by the evolutionary history of the binary system in which they lie. 
For instance, pulsars with low-mass WD companions have gone through a long-lasting LMXB phase which has likely increased 
their mass anywhere from $\sim\!0.1-0.5\,{\rm M}_\odot$ (see the previous Section~\ref{sec:binaries} on pulsar recycling). 
Double NS binaries, on the other hand, should reflect the birth mass more closely since the first-born NS only accretes $<0.01\,M_\odot$ 
during/following a common-envelope evolution \citep{tlk12} while the second NS should have accreted none. 
Comparing the masses from the two sub-populations might therefore provide a powerful probe for mass transfer in binaries.

\subsection{The contribution of SKA}
SKA1 is expected to increase the number of known pulsar binaries by more than a factor of five, which should therefore grow 
the NS mass sample size by a similar factor. This naive assumption is certainly justified for relativistic NS--NS binaries since their mass 
measurements rely on radio timing only, while in other binaries they suffer the caveat that they require optical data as well. 
Projects such as the Thirty-Meter Telescope and the European Extremely Large Telescope will provide a ten-fold increase in sensitivity, 
thus enabling to push down the luminosity function of the probed systems to sustain the growth of mass measurements in these systems as well.

Given a factor of ten increase in the mass sample with SKA2, how well can we expect to determine the minimum and maximum mass range of NSs? 
Based on the distribution of \citet{kkdt13}, i.e. a normal distribution ${\cal N}(\mu=1.33\;M_{\odot}, \sigma=0.12\;M_{\odot})$, the probability of finding a 
mass below the putative 1.08\,M$_\odot$ lower limit is 1.86\%, which implies that a 200-NS masses sample from NS--NS systems would 
have a 97.7\% chance of containing such a low-mass NS (should they exist). If a low-mass limit cutoff exists, it will as well manifest as a 
skewness in the distribution, which would lack data at the low end. 
Regarding the upper mass limit, we turn toward NS--WD systems. Here, for simplicity, we assume that they have the same underlying 
birth-mass distribution as the NS--NS systems (i.e. ${\cal N}(\mu=1.33, \sigma=0.12)$), and have then accreted some arbitrary mass 
that we model using a gamma distribution with a scale parameter 0.15 and a shape parameter 2.0. 
Based on this we infer the probability of finding a mass above 2.4\,M$_\odot$ is 0.42\%, thus corresponding to 47\% odds of 
having a NS above that limit in a sample of 150 (should such a NS exist). 
From the distribution of NS masses and eccentricities it is even possible to place limits on the
gravitational binding energy of NSs \citep{pdl+05}, released via neutrinos during the SN event. Thus future SKA observations can also via
this method help to narrow down the EoS of dense nuclear matter \citep[see the chapter by][]{wxe+14}.

\section{Neutron Star Kicks and Velocities}

Soon after the discovery of pulsars, it was recognized that their average space velocity greatly exceeds
that of their progenitor stars \citep{go70}. At the present time, with proper motions measured for hundreds
of pulsars via timing or VLBI, it is clear that most radio pulsars
receive a large kick ($\sim\!400$~km~s$^{-1}$) during their formation event \citep{hllk05}.
However, a number of plausible processes remain for the exact physical mechanism responsible 
for imparting this kick during the SN \citep{lcc01,jan12}.  
Many pulsars are observed in globular clusters which have small escape velocities
of $\lesssim 40\;{\rm km}\,{\rm s}^{-1}$. 
Theoretical studies of electron capture SNe and the accretion-induced collapse (AIC) of massive WDs 
provide a plausible NS formation mechanism which predicts a 
low kick velocity \citep{plp+04,spr10,tsyl13}. In addition, there is the interesting question of a possible alignment 
at birth between the pulsar velocity vector and its spin axis, as suggested by recent observational evidence and modelling \citep{nsk+13}. 

In order to differentiate between the competing models for NS birth kicks, reliable
velocity estimates are needed for a larger fraction of the known pulsar population. The fact that the 
radial component of velocity is inaccessible for pulsars in all but a very few cases sharpens the
need for a large sample of objects. As of early 2014, transverse velocity estimates are available for 
around 270 pulsars\footnote{http://www.atnf.csiro.au/people/pulsar/psrcat/}, but most of these
are of low precision, either because of the proper motion measurement itself or (more commonly)
uncertainties in the pulsar distance  
provided from the pulsar dispersion measure and the NE2001 model of the Galactic electron density distribution \citep{cl02,dtbr09,dbc+11}.
In order to improve the reliability and size of the pulsar velocity distribution sample, three things are needed:
i) Additional proper motion measurements, via timing or high resolution astrometry;
ii) Accurate, model-independent distance measurements for a larger fraction of the objects with proper
motion measurements (again via timing and - more commonly - high resolution astrometry); and
iii) A better model of the Galactic electron density distribution, allowing more reliable model-dependent
distance estimates for the remaining pulsars.

\subsection{The contribution of SKA}
With its phenomenal sensitivity, SKA will be able to measure proper motions and parallax distances
for radio pulsars via the current approaches (high precision timing and high resolution astrometric
observations) but with greatly improved accuracy, and for a much larger number of systems \citep{stw+11}.
For recycled pulsars with very stable rotation properties, SKA will be able to measure proper motion 
and often parallax via timing.  In many cases, this will come as a byproduct of programs undertaken 
for other purposes, as described earlier in this chapter.
The remaining pulsars can be addressed via astrometry. With the moderate baselines of SKA1-MID 
($\sim\!100$~km) it will be possible to measure positions accurate to several milli-arcseconds in a single 
observation for all but the faintest pulsars.  This will be sufficient to measure a proper motion for
the majority of pulsars if several observations are made over a multi-year period.  

However, to obtain the parallax measurements necessary to anchor the distance scale, 
much higher angular resolution will be needed. This can be 
achieved using a phased SKA1-MID and SKA1-SUR as part of a wider intercontinental VLBI network.

The biggest impact of SKA2 for pulsar velocity studies is the extension of baselines to SKA1-MID. With
a resolution of a few tens of milli-arcseconds or better at GHz frequencies, coupled with 10$\times$ higher
sensitivity than SKA1-MID, SKA2 will be able to perform ultra-precision astrometry on the 
majority of the entire radio pulsar population visible from the site. In terms of providing the accurate 
velocity and distance information needed to characterise the pulsar population, SKA2 as currently
envisaged encompasses all of the required capabilities.

\section{A brief summary of the impact of early phase SKA1 (50\% SKA1)}

We end this chapter with a brief summary of the outlooks for early deployment of SKA1 and its expected impact on 
pulsar science \citep[see also][]{ks14}.

Early deployment of SKA1 at 50\% sensitivity would not provide access to the faintest pulsars of the population.
Simulations for SKA1 by \citet{kbk+14} predict detection of $\sim10\,000$ normal pulsars and $\sim 1500$ MSPs, whereas 
for early phase SKA1 (50\% SKA1) the expected numbers are roughly 6000 and 750, respectively (keeping the integration time the same).
To partly compensate for the sensitivity loss one could, for example, double the integration time.
In that case the predicted number of detections are about 7500 normal pulsars and 950 MSPs. However, an increase in integration time would
also largely increase the required amount of data processing given that the latter scales with the integration time cubed.  
An additional handicap for discovering the fast spinning MSPs (as well as double NS and NS--BH binaries) 
would be the increase in Doppler-smearing of the radio pulses
using longer integration times, given that many MSPs are being highly accelerated in tight binaries.

The expected sensitivity limit for the SKA1-MID survey can be compared with, for example, the HTRU survey \citep{kjv+10} ongoing at the Parkes Radio Telescope
and the Arecibo PALFA survey \citep{cfl+06}. Whereas the sensitivity for SKA1~50\% would only be about half the sensitivity
of the Arecibo PALFA survey (which, on the other hand, has a very limited sky coverage), it would still be significantly better than the 
HTRU survey by a factor of 2--6, depending on the Galactic latitude.
Another comparison can be made to the complete 64~dishes MeerKAT Radio Telescope. MeerKAT will be operational in 2017 
and will later become an integrated part of SKA1-MID. Early deployment of SKA1 at 50\% would still have a sensitivity about two times better 
than that of MeerKAT.

For the measurement of NS kicks and velocities, early-phase observations with SKA1 at 50\% sensitivity would be unable to access the faintest pulsars, 
limiting the maximum number of accurate transverse velocity measurements that could be made (whether by timing with SKA1-MID, imaging with SKA1-MID, 
or imaging with SKA1-VLBI). The pool of remaining targets still represents a factor-of-many increase over the current state of affairs, 
and in any case is likely larger than the observing time available in early-phase SKA1.

As pointed out elsewhere, the SKA1 is not only a simple stepping stone
towards SKA2. Due to limitations in processing power, it is
unlikely that the full area of the completed SKA can be utilized for a
blind, large-scale survey for some time to come. 
SKA1 with a highly concentrated core represents a significant fraction of
collecting area usable for surveys with SKA2.
Although the outcome will depend on the number of dishes, the bandwidth
and the central frequency of a given survey (or a composite survey covered by SKA1-LOW and SKA1-MID), the total number of beams, 
and the computer power available,
significant achievements can even be made in finding MSPs and binary pulsars 
during early deployment pulsar searching at the 50\% SKA~level.

\bibliographystyle{apj}
\bibliography{tauris_refs}

\end{document}